\documentclass[letterpaper, 12 pt]{article}
\usepackage[a4paper, total={6in, 8in}]{geometry}
\usepackage{graphicx} 
\usepackage{amsmath}
\usepackage{graphicx}
\usepackage{xcolor}
\usepackage{cancel}
\usepackage{lipsum}
\usepackage{cite}
\usepackage{amsmath,amssymb}
\usepackage{physics}

\title{Relativistic generalised uncertainty and the corrected vacuum energy}
\author{ Akshat Pandey\footnote{apandey.physics@gmail.com} \\ \normalsize Department of Physics, Shiv Nadar Institution of Eminence \\ \normalsize Greater Noida, Uttar Pradesh-201314, India.}

\date{}

\begin{document}

\maketitle

\begin{abstract}

In this short paper we make use of the recent covariant extension of the generalised uncertainty principle to study the corrections to the vacuum energy of the simplest scalar quantum field theory. We then calculate the modifications to the Casimir effect that such a correction to the vacuum energy would bring about. We emphasise that these corrections are indeed physical.

\end{abstract}

\section{Introduction}

Predictions of a minimum measurable length have been a common feature of theories of Quantum gravity (QG) \cite{sabine1, sabine2, sabine3, sprenger, minic}. Within QG phenomenology, the Generalised Uncertainty Principle (GUP) which extends the Heisenberg Uncertainty Principle to include high momenta cut-offs has been an important approach at incorporating the idea of the smallest measurable length. \cite{kempf, das1, das2, das3, G1, G2, G3, G4, G5, G6, G7, G8, G9, G10}. Recently, there have been attempts to put GUP into a covariant framework \cite{G11, RGUP, CGUP, CGUPQFT}. In particular, Torodinnov et al extended GUP to relativistic scenarios \cite{RGUP} with the modified commutation relation taking the form

\begin{equation}
    [x^\mu, p^\nu]= i \hbar (1 - \gamma p^{\alpha}p_{\alpha}) \eta ^{\mu \nu} + i \hbar {\gamma} p^\mu p^\nu
\end{equation}

Here $c=1$, and all Greek indices $\in \{0,1,2,3\}$. $\gamma$ is the GUP parameter with dimensions of inverse momentum squared, $\gamma = \frac{\gamma_0}{M_{Pl}^2}$, $\gamma_0$ is a dimensionless that fixes the GUP scale. 
The Relativistic GUP (RGUP) thus points at a minimum  position uncertainty that is Lorentz invariant. This ensures that Generalised uncertainty can be formulated in the context of Quantum Field theory, making it possible to calculate minimal length induced corrections to well-known effects including, for example, scatterring amplitudes and electron-muon cross-sections \cite{GUPQFT1, GUPQFT2, GUPG}. However, there seems to be no work exploring the measurable effects of RGUP to free field theories. 

The purpose of this present work is twofold. Firstly we show how even for the simplest case of a free scalar field, imposing RGUP leads to modification in the vacuum energy of the field. Secondly, by calculating the vacuum fluctuations and the corrected Casimir force, we emphasise that this shift in the vacuum energy is indeed physical and can have measurable consequences.

This paper is organized as follows. In section 2 we review RGUP modifications to a scalar field and then go on to calculate the vacuum energy. In section 3, using the results for the vacuum energy we study fluctuations and calculated the corrected Casimir force. We end with a summary in section 4. For the rest of the paper we assume natural units, that is, $\hbar=c=1$.

\section{Modified Scalar field theory and the Vacuum}

To start, we notice that the commutation relations (1) imply that the variables $x^{\mu}$ and $p^\nu$ are not canonically conjugate. We can thus define additional variables $x_{0}^\mu$ and $p_{0}^\nu$ that obey the standard commutation rules 

\begin{equation}
    [x_0^\mu, p_0^\nu]= i \eta ^{\mu \nu}
\end{equation}

Further, we can see that $p_0^\mu$ are related to $p^\mu$ via

\begin{equation}
    p_{\mu}= p_{0 \mu}(1 + \gamma p_0^\mu p_{0 \mu})
\end{equation}

Thus a field $\phi$ that obeys the standard Klein-Gordon equation for $p^\mu$ could thus be rewritten as

\begin{equation}
    (\Box + 2 \gamma \Box ^2 +m^2) \phi = 0
\end{equation}

where $\Box = \partial_\mu \partial^\mu$. Starting with (3), we restrict to terms of the order $\gamma$ and neglect terms $\mathcal{O}(\gamma ^2)$ .The $\gamma$ term above and in all subsequent equations is interpreted as RGUP based Quantum gravity corrections to Quantum field theories \cite{das1}. For the above equation the Lagrangian density as obtained by Bosso et al \cite{GUPQFT1} is 
\begin{equation}
    \mathcal{L}= \frac{1}{2}\partial^\mu \phi \partial_{\mu} \phi - \frac{1}{2}m^2 \phi ^2 + \gamma \partial^\nu \partial_{\nu} \partial^\mu \phi \partial_{\mu} \phi
\end{equation}

This additional term in the Lagrangian density can be thought of as another degree of freedom $\partial_\mu \partial^\mu \phi $ at each point in space in addition to $\phi$

Defining the conjugate momenta 
\begin{equation}
    \begin{aligned} \pi_\alpha^\alpha & =\frac{\partial \mathcal{L}}{\partial\left(\partial_\alpha \partial^\alpha \dot{\phi}\right)}=2 \gamma \dot{\phi} \\ \pi & =\frac{\partial \mathcal{L}}{\partial(\dot{\phi})}-\partial_\beta \partial^\beta \pi_\alpha^\alpha=\dot{\phi}\end{aligned}
\end{equation}

Using (5) and (6), the Hamiltonian density can be calculated
\begin{equation}
   \mathcal{H} =\frac{1}{2} \pi ^2+\frac{1}{2} m \phi^2+\frac{1}{2}(\nabla \phi)^2+\gamma(\nabla \phi) \cdot\left(\partial_\sigma \partial^\sigma \nabla \phi\right).
\end{equation}

The Hamiltonian density is that of a free scalar field with an extra $\gamma$ term. Here $\gamma \sim 10^{-38}$ ensuring that the theory is Ostragradsky stable \cite{thesis}. Alternatively the De Donder Weyl formulation \cite {de} could have been used to yield a much simpler looking Hamiltonian density where the $\gamma$ terms would cancel out and the RGUP corrections be encoded in the disperson relation and the field definition. However, we shall stick to equation (7) as our Hamiltonian density. This makes calculations simpler as written this way, we can use the standard relativistic dispersion relations and the following standard definition for a real scalar field

\begin{equation}
    \phi(\vec{x}, t)=\int \frac{d^3 k}{(2 \pi)^3} \frac{1}{\sqrt{2 \omega_{\vec{k}}}}\left(a_{\vec{k}} e^{-i(\omega t - \vec{k} \cdot \vec{x})}+a_{\vec{k}}^{\dagger} e^{+i (\omega t - \vec{k} \cdot \vec{x})}\right).
\end{equation} 

Note, $\vec{k}$ is simply $\vec{p}_0$ in natural units and thus obeys the standard dispersion relations.  
Computing terms in the Hamiltonian density using the above field definition is straightforward.

Now we wish to calculate the vacuum energy which turns out to be

\begin{equation}
 \mel{0}{H}{0} = \int d^3x \int \frac{d^3k}{(2 \pi)^3} \frac{1}{2 \omega_{\vec{k}}}\frac{1}{2} \left( \left(\omega_{\vec{k}}^2 + \vec{k}^2 + m^2   \right) - \gamma \left(\omega_{\vec{k}}^2(\vec{k}^2 + m^2) \right) + (\vec{k}^2 + m^2)^2  \right)
\end{equation}

Plugging in the dispersion relations, we end up with

\begin{equation}
    \mel{0}{H}{0} = \int d^3x \int \frac{d^3k}{(2 \pi)^3} \frac{1}{2}(\omega_{\vec{k}} - \gamma \omega_{\vec{k}}^3)
\end{equation}

The first term in the vacuum density is simply the ground state energy of the harmonic oscillator integrated over all momentum modes, which is the standard result. The second is the correction due to RGUP. While the modifications to the commutation relations and the Hamiltonian were fairly involved, the $\gamma$ modification to the vacuum energy has a rather simple form. This simplicity lets us sketch some predictions for measurements of these Quantum gravity corrections. In the following section we see this in the context of the Casimir effect.  

\section{Corrections to the Casimir Force}

The Casimir force is a physical consequence of the non-trivial vacuum in QFT, experienced by the bounding plates when a Quantum field is restricted to a finite region \cite{Casimir}. GUP corrections to the Casimir force have been previously worked out by \cite{CasGUP}. Another interesting paper is by Blasone et al. where they use GUP arguements to put forward a heuristic derivation of the Casimir effect \cite{CasGUP2}. 

This section is however different from the above mentioned articles as we are making use of RGUP and working in a field theoretic framework. Our goal is not to give a careful solid state physics based account of the Casimir force in a laboratory setup, but to simply illustrate that the modifications to the vacuum that are brought about by Generalised Uncertainty can have measurable consequences. Further, for simplicity of calculations we will restrict our analysis to 1+1 dimensions and to a massless scalar field \cite{Zee}.

The vacuum energy in 1+1 dimensions is
\begin{equation}
    \mel{0}{H}{0} = \int dx \int \frac{dk}{(2 \pi)} \frac{1}{2}(\omega_{k} - \gamma \omega_{k}^3)
\end{equation}

We impose periodic boundary conditions by restricting the field in a box of length $L$. Further, within this box, we calculate the vacuum energy $E(l)$ between two thin metal sheets seperated by a distance $l$ such that the total energy would be $E(l) + E(L-l)$. Following (7) $E(l)$ this turns out to be 

\begin{equation}
    E(l)= \sum^k \frac{1}{2}\omega_k - \sum^k \frac{1}{2} \gamma \omega_k^3
\end{equation}

Making use of the mode expansion for the energy $\omega= \frac{n \pi}{l}$ in the box We replace this infinite sum over $k$ with sum over all $n$. Thus, the energy becomes
\begin{equation}
    E(l)= \frac{\pi}{2l} \sum^n n - \frac{\gamma \pi ^3}{2 l^3} \sum^n n^3
\end{equation}  

We make use of exponential regularisation to obtain finite results from these infinite sums. Physically this corresponds to introducing a UV cutoff in terms of a length scale $\alpha$, that is to say that the higher frequency modes contribute less and less to the sum. 

Equation (10) thus becomes

\begin{equation}
\begin{aligned}
     E(l) &= \frac{\pi}{2l} \sum_{n=1}^\infty n e^{-\alpha n/l} - \frac{\gamma \pi ^3}{2 l^3} \sum_{n=1}^\infty n^3 e^{-\alpha n /l}\\&= -\frac{\pi}{2} \frac{\partial}{\partial \alpha} \sum_{n=1}^\infty  e^{-\alpha n/l} + \frac{\gamma \pi ^3}{2 }\frac{\partial^3}{\partial \alpha^3} \sum_{n=1}^\infty  e^{-\alpha n /l}\\& = \frac{\pi}{2}\frac{\partial}{\partial \alpha}\left(\frac{e^{-\alpha/l}}{e^{-\alpha/l}-1}  \right) - \frac{\gamma \pi^3}{2}\frac{\partial^2}{\partial \alpha ^2}\left(\frac{\partial}{\partial \alpha}\left(\frac{e^{-\alpha/l}}{e^{-\alpha/l}-1}  \right) \right)\\& = \frac{\pi}{2l}\frac{e^{-\alpha/l}}{(e^{-\alpha/l}-1)^2}  - \frac{\gamma \pi^3}{2}\frac{\partial^2}{\partial \alpha ^2} \left( \frac{e^{-\alpha/l}}{(e^{-\alpha/l}-1)^2}  \right)
\end{aligned}
\end{equation}

The exponential suppression enables us to Taylor expand about $\frac{\alpha}{l} \approx 0$. We get

\begin{equation}
\begin{aligned}
    E(l)&= \frac{\pi}{2l}\left(\frac{l^2}{\alpha^2} - \frac{1}{12} + \frac{\alpha^2}{240 l^2} + ...\right) - \frac{\gamma \pi^3}{2l}\frac{\partial^2}{\partial \alpha^2}\left(\frac{l^2}{\alpha^2} - \frac{1}{12} + \frac{\alpha^2}{240 l^2} + ...\right) \\ &= \frac{\pi l}{2\alpha^2} - \frac{\pi}{24 l} + \frac{\gamma \pi^3 l}{12 \alpha^4} - \frac{\gamma \pi^3}{240 l^3} + \mathcal{O}(\alpha^2)    
 \end{aligned}   
\end{equation}

The energy for the rest of the system can be calculated by substituting $l \xrightarrow{}L-l$. The total energy turns out to be

\begin{equation}
\begin{aligned}
    E &= E(l) + E(L-l) \\ &= \frac{\pi L}{2 \alpha^2} - \frac{\pi}{24}\left( \frac{1}{l} + \frac{1}{L-l} \right) + \frac{\gamma \pi^3 L}{12 \alpha^4} - \frac{\gamma \pi^3}{240}\left(\frac{1}{l^3} + \frac{1}{(L-l)^3} \right) + \mathcal{O}(\alpha^2)
\end{aligned}
\end{equation} 

Now, we can write down the Casimir force

\begin{equation}
    F_{Casimir} = -\frac{d E}{d l} = -\frac{\pi}{24 l^2} - \frac{\gamma \pi^3}{720 l^4} + \mathcal{F}(\alpha/l , l/L)
\end{equation}

$\mathcal{F}$ represents other terms that show up in the force. These terms are dependent on the IR and UV regulators $L$ and $\alpha$, respectively. These terms vanish as we take $L \xrightarrow{} \infty$ and $\alpha \xrightarrow{}0$, and can simply be ignored.

We thus end up with an expression for the corrected Casimir force. Note that upon taking the RGUP parameter $\gamma \xrightarrow{}0$, we recover the standard expression for the Casimir force. The sign of the $\gamma$ term would determine whether RGUP strengthens or weakens the attractive force between the bounding plates. Further, the correction term in the force goes as $l^{-4}$ implying that the correction is relevant only for very small $l$ and is quickly suppressed when the distance between the plates is increased. Note also that this description cannot be valid for arbitrary small $l$ as the very premise of the calculation involved a built-in minimal length. Thus, we assume $l$ to be much larger than the minimum length scale.

Alternatively, starting from equation (13), we could have crudely made the identification $\sum_n n \sim \frac{-1}{12}$ and $\sum_n n^3 \sim \frac{1}{120}$. However, it is instructive to use exponential regularization as it also offers physical explanation for the finiteness of the final answer, namely the UV cutoff \cite{Zee}. Also, though we chose exponential regularization for its usefulness in analytic calculations, other regulators like the zeta-function regulator could have also been used. For an interesting discussion on regulator choices in QFT and its connections to number theory, see \cite{tony}.

Note also that our analysis was only for a toy scalar model. To extend this to a situation which is physically testable would require a similar analysis for the electromagnetic field in 3+1 dimensions. In addition to this, the infinitesimal thickness condition on the conducting plates will have to be relaxed and the response of these plates to the electromagnetic field will also have to be taken into account. Apart from this, we believe, a quantitative analysis of how large the corrections would be given the smallness of the parameters $\gamma$ and $l$, also deserves further investigation.
 
\section{Summary}

In this paper we studied some consequences of the Generalised Uncertainty Principle applied to the simplest Relativistic Quantum field Theory. In particular, we worked out how the vacuum energy gets modified by imposing a Covariant Generalised Uncertainty Relation on the field. We emphasised that these modification to the vacuum and vacuum fluctuations are indeed real and to this effect calculated the corrections to the Casimir force that such a modification of the vacuum energy would lead to.

\section*{Acknowledgements}
I thank the anonymous referee for their comments on an earlier version of this manuscript. I also wish to thank Noel Jobu and Rahul Ghosh for useful discussions.

\end{document}